# A simplified model for expected development of the SARS-CoV-2 (Corona) spread in Germany and US after social distancing

Franz-Josef Schmitt

Martin-Luther-Universität Halle-Wittenberg, Institut für Physik, 06122 Halle

Email: franz-josef.schmitt@physik.uni-halle.de

**Abstract**

Widespread opinions and discussion exist regarding the efficiency of social distancing after crucial spread of the SARS-CoV-2 virus during the actual Covid-19 pandemic. While Germany has released a federal law that prohibits any type of direct contact for more than 2 people other countries including the US released curfews.

People are now wondering whether these measures are helpful to stop or hamper the Covid-19 pandemic and to limit the spread of the new corona virus.

A quantitative statement on this question depends on many parameters that are difficult to grasp mathematically and cannot therefore be made conclusively (they include consistent adherence to the measures decided, the estimated number of unreported cases, the possible limitation by test capacities, possible mutations of the virus, etc ...). However, it turns out that a reduction in the actual daily new infection rate (actual daily growth rate of reported cases, in short: infection rate) from the current value of 30-35% in the US to 10% would be extremely effective in stopping the spread of the virus. The severe restrictions in Germany which closed any public events, schools and universities a week ago might already have contributed to a reduction of the growth rate of reported cases below 30%.

It must be clearly stated that a reduction in the infection rate from 30% to e.g. 15% in no way means that it would be sufficient to halve social contacts. Since the infection routes are not known in detail, it is rather likely that such a significant reduction in the infection rate can only be achieved if social contacts are completely avoided for a limited period of time, because in addition to personal contact, there are other infection routes via e.g. smear infections (e.g. in shops or anywhere) that cannot be easily influenced by social distancing.

Furthermore, success cannot be assumed if the new infection rate drops to a value of 10% or less within a few days. It has to continuously approach zero, over a period of about 14 days, to guarantee success. Therefore it is of significant importance to keep social distancing up until a significant drop of the accumulated number of all active cases is reported and authorities agree for abolition of the ban.

Nevertheless, a simplified model presented here can give a glimpse for correlation between the infection rate and the expected growth of active infections and show that it is possible to even reduce the number of infected cases respecting actual properties of SARS-CoV-2 reported in the literature

At the request of some friends, colleagues in this study I would like to contribute to the actual discussion with a simple, understandable model that can be communicated well in its manageable form, despite the aforementioned imponderables. The key messages of the highly simplified study presented here are:

i) Social distancing, if it is followed consistently, is very successful and suitable to interrupt the spread of the SARS-CoV-2 virus.
ii) While the measures in the US seem to be behind their efficiency as compared to Germany a consequent compliance is suitable to interrupt Covid-19 in the US with comparable efficiency.
iii) For the end of March 2020, the model shows a significant decrease in the number of registered new infections compared to an unchecked spread without social distancing. From these numbers the efficiency of all measures can be judged and controlled.
iv) If the measures are successful, the model will be followed by maxima with an expected low (in Germany) or medium (in the US) six-digit number of active infections between April 1 and 20, 2020.
v) A comparatively small difference in the reduced new infection rate achieved through social distancing can mean a multiple of infections at the peak of the pandemic

- vi) The actual development respected up to March 23 2020 leads to simulation results from the simple model presented that a reduction of the infection rate to 20% culminates in a maximum of 350.000 active infections in mid April while a strict reduction to 10% might even stop the active infections at 80.000.
- vii) In the US an immediate reduction to 20% can stop the registered infections at a maximum of 1.7 million around April 20$^{th}$ while 300.000 cases are expected when the infection rate can be reduced to 10%. Therefore the small advance of the measures in Germany leads to a rise of the expected registered infection by a factor of 4-5 if the same measures are applied from now.
- viii) Delays in implementation lead to a significant increase in the maximum number of infections. This increase corresponds approximately to the current new infection rate in the US (approx. 34%), applied to the maximum number of infections that occur. This means that a maximum of 1 million infected people will rise to 1.34 million infected people if the consistent implementation of the measures is delayed by one day. So a single day's delay possibly means several hundred thousand additional infections.

The contact ban is effective, but the dependence on time and on the initial and boundary conditions is strong.

Main aim of the study presented here is to alleviate the fear of the virus but clarify and insist on the need of strict measures and assessment. If we strictly follow the rules, I am convinced, we will be able to interrupt the spread of Covid-19 within a week or two of uncertainty on the timeline.
I now ask all readers to attach the trust they have gained to the basic condition of interrupting their interpersonal contacts until the ban is canceled by the authorities. The study shows that we can control the SARS-CoV-2 virus in exactly such way.

A model for the spread of the SARS-CoV-2 virus using a set of differential equations can be e.g. found in Ref. [1-4]. However, it was not the aim of this study to find a model that was as precise as possible, but to give one intuitive simple equation that enables us to get a feeling for what´s going on and provide an estimate that corresponds qualitatively to the courses everyone can model on the website [4].

**Introduction**

The literature contains studies that calculate and visualize the effectiveness of measures known as social distancing [1-3]. Sources are available on the Internet that calculate the spread of the SARS-CoV-2 virus in real time with individually adjustable parameters for incubation time, infection rate, death rate and recovery time [4].

Most of these studies have the following in common: i) They show that social distancing can significantly slow down the spread of the virus [1-3]. ii) They show that the effectiveness depends very much on the initial and boundary conditions of social distancing and the results strongly vary by orders of magnitude if the process is initiated a few days earlier or later or the effectiveness of the measure is only slightly higher or lower [4].
iii) The details of the studies are difficult to understand for many readers and therefore a widespread trust on the efficiency on the actual measures regarding the Covid-19 pandemic can not easily be derived from scientific literature.

Studies such as [3] show that the effectiveness of social distancing is particularly great for pandemics with a basic reproductive number of $R_0$ = 1.5 .. 2.5. For the Covid-19 pandemic caused by the SARS-CoV-2 virus, a basic reproduction number $R_0$ = 2.6 is estimated, with an uncertainty range from 1.5 to 3.5 [5]. The new infection rate of 28% (in Germany) and 34% (in the US) for 10 days of infectiousness, which is the basis for this study, corresponds approximately to a basic reproductive number $R_0$ = 2.8 .. 3.4.

$R_0$ specifies how many new cases are caused by an existing case on average. With $R_0$ > 1, the virus spreads exponentially at times, but for $R_0$ <1 the spread stops on its own.

A large number of studies have now calculated the number of basic reproductions, the incubation period and the duration of the infectiousness based on the number of cases from China. The results of this study were taken into account in the underlying model as part of the simplification [5-15] (for details see the following section: Mathematical Model).

The aim of the presentation presented here is now in no way to challenge existing studies on the effectiveness of social distancing, but to demonstrate the effectiveness of the measure. The basic principles of the growth curve of the SARS-CoV-2 virus should be visualized with the simplest possible modeling. In addition, the essential key points of the simulation result should be made clear, namely the effectiveness of the measure in relation to the start time and the possible achievable reduction of the infected.

The basis is therefore not the basic number of reproductions but the current daily rise of registered infections (in short: infection rate). The actual infection rate is calculated by averaging the growth of the observed new infections per day for a period of the last 10 days. In contrast to the basic reproductive rate, the infection rate can be read directly from the published figures for Covid-19. It was about 30% in Germany until March 21, 2020, averaged over the increase in new infections since March 1, 2020 but started to reduce significantly from March 21 to a value between 15-20 % for the last days [4]. This clear reduction is a very promising observation regarding the forthcoming spread as it will be pointed out in the following.

**Mathematical model**

The following simplifying assumptions are made to construct the mathematical background:

1. The current rate of new infections is determined as the mean of the increase in registered infected people over the period March 10 – March 23 according to the data published in [16].

2. Until March 23 the data published under [16] for the total number of registered infected people will be used in the model. From March 24 a numerical simulation of the further course will take place.

3. The current measures of social distancing lead to an as yet unknown reduction of the infection rate, which (in our model) will only become fully noticeable by end of March 2020. The infection rates k shown in Fig. 1 for Germany and Fig. 2 for the US are approximately the actual infection rates that can only been seen as clearly reduced growth of infections by March 29 as SARS-CoV-2 has an incubation time of 5-6 days [5-15].

4. Since events have been banned for some time and schools have been closed, it is assumed that the current infection rate lies in the middle between the infection rate achieved by social distancing and the original rate of up to 34% as calculated for the US.

5. In the model it is assumed that newly infected people are themselves infectious for 10 days. This value should be assumed to be rather long, since values in the literature range from 3 days to a maximum of 10 days [5-15].

6. It is assumed that newly infected people fall out of the statistics after 21 days because they have either recovered or died [5-15].

If we denote the unknown current infection rate by *k*, then during the growth phase of the pandemic the number of active registered infected people on day *n + 1* calculates to:

1. $$I(n+1) = I(n) + (I(n) - I(n-10)) \cdot k - I(n-21)$$

On the right side, the first term denotes the infected people of day *n*, the second term adds the people who have been registered in the last 10 days multiplied with the actual infection rate *k*. This number is reduced by the number of infected people registered 21 days ago (third term), since they have now either recovered or died.

At this point, two further simplifications should be mentioned:

1. The model does not take into account the unreported number of infected people who are not registered. It is therefore assumed that the number of registered infected people forms a representative group for the actually real infected and is not limited by the test capacities or a variable test behavior or other factors.
2. The model does not take into account quarantine. The registered infected people are considered to contribute to the further infections. In act one could suppose that it is right opposite because these people are in quarantine, but due to the number of unreported cases it can be assumed that their number is proportional to the infected people who are not in quarantine actually. So they form a representative number that contributes to further growth of infections via the infection rate *k*.

**Results and Discussion**

Fig. 1 shows the course of *I (n)* according to the given equation for Germany as expected active infections to be in the range from March 15 to April 30 for various infection rates *k*, possibly resulting from social distancing. An initial conservative infection rate of 31% was summed according to the data presented in [16]. It should be noted that the values reported are always subject to statistical fluctuations and therefore do not correspond exactly to the "real" daily values but need to be averaged over a certain time span.

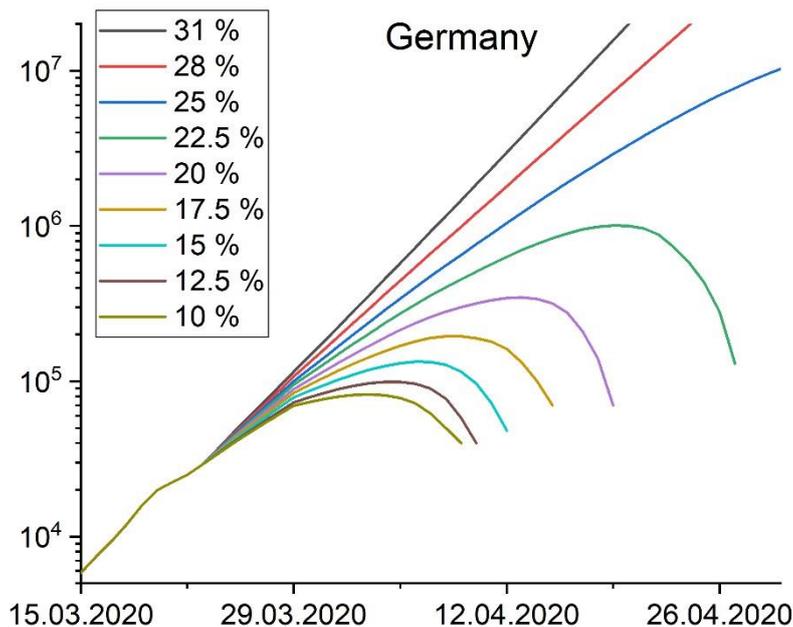

**Fig. 1 Course of *I (n)* in logarithmic representation, calculated for Germany from March 24 with actual real infection rate of *k = 0.1* to *k = 0.31* as it will be noticed in corresponding low growth rates by March 29. Until March 23 the infections reported under [16] were used.**

Fig. 2 shows the exactly same simulation from March 1st for the US the only differences are:
1. The actual growth rate in the US is determined to be slightly higher with 0.345 rather than 0.28 .. 0.31 as determined for Germany
2. The actual number of reported infected cases are slightly higher with 43734 registered infections on March 23 in the US as compared to 29056 registered infections on March 23 in Germany

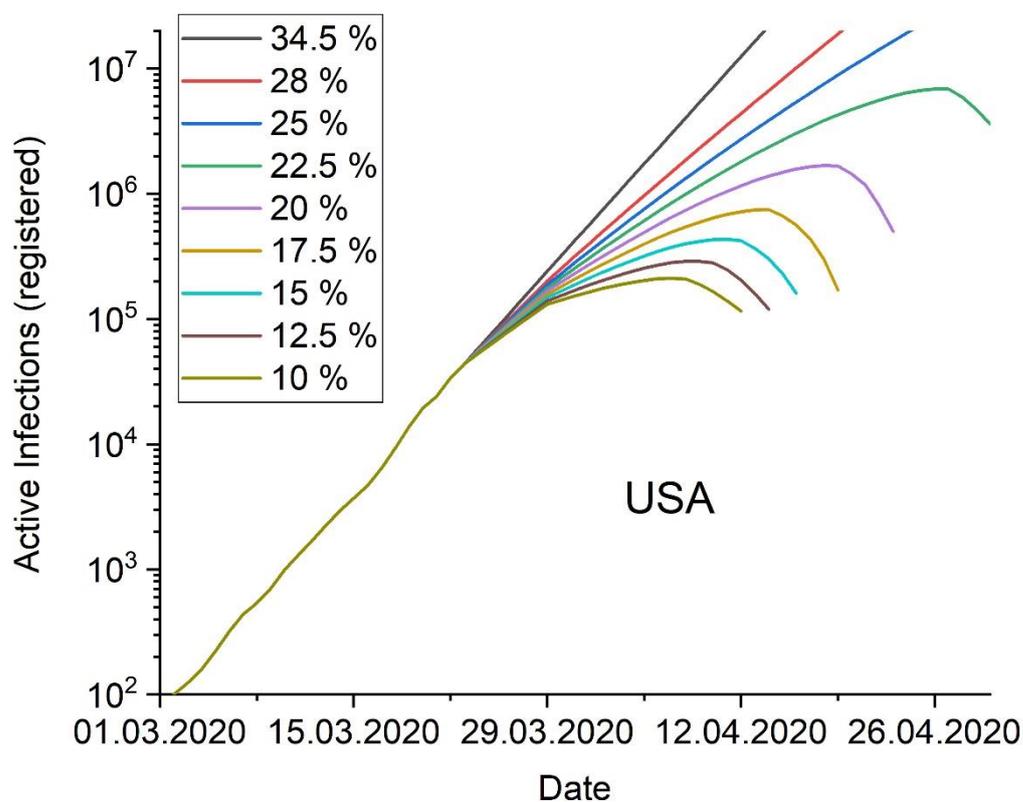

**Fig. 2** Course of *I (n)* in logarithmic representation, calculated for the US from March 24 with actual real infection rate of *k = 0.1* to *k = 0.345* as it will be noticed in corresponding low growth rates by March 29. Until March 23 the infections reported under [4] were used.

The simulation result shows the following properties of the calculated course:
  i)   Without measures, an exponential increase in new infections continues until natural restrictions curb the further spread of the virus which will possibly occur after passing 10 million infections (in Fig. 1 and Fig. 2 an apparent linear increase as a logarithmic representation, linear representation see Fig. 3).
  ii)  By reducing the new infection rate, it can be achieved that the number of active infections reaches a maximum and then decreases (the drop in the curve requires a modification of the equation and is therefore not modelled here).
  iii) The reduction in the new infection rate must be sufficiently strong to lead to a maximum and subsequent decrease in new infections (initially 22.5% or less and then continuously decreasing).
  iv)  The effectiveness of the measures will probably be significantly reflected in the number of new infections at the end of March.
  v)   Falling numbers of actively infected people are (depending on the scenario) not to be expected before the first week of April, but more likely from mid April with several days delay in the US as compared to Germany.
  vi)  The actual development respected up to March 23 2020 leads to simulation results from the simple model presented that a reduction of the infection rate to 20% culminates in a maximum of 350.000 active infections in mid April while a strict reduction to 10% might even stop the active infections at 80.000.
  vii) In the US an immediate reduction to 20% can stop the registered infections at a maximum of 1.7 million around April 20[th] while 300.000 cases are expected when the infection rate can be reduced to 10%.
  viii) The maximum is run through the later, the higher the new infection rate (between April 3 and April 7 for 12.5% and Germany and US, respectively and April 20 and 27 for 22.5% and Germany and US, respectively, see Fig. 2 and 3)

- ix) The different course of the new infection rates in the next few days provides information about the success of the school closings and event cancellations as well as the public awareness before the decision to ban the contact (see Fig. 4).
- x) From March 29, the decision to ban contact will also be reflected in the new infection rate and deliver clear information if further steps are necessary
- xi) It will be clear at the end of March which scenario we can expect and whether the measures taken are sufficient to interrupt the spread (see Fig. 4).
- xii) the decrease in infections after passing through the maximum depends on many other factors which are not adequately described by the model shown. The discussion of the decline in detail is therefore not the subject of the current study.

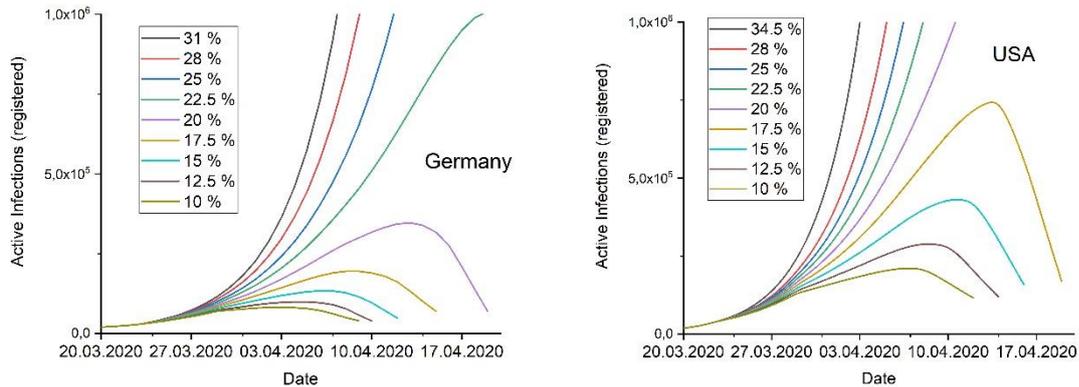

**Fig. 3 Enlargement of the period March 20, 2020 - April 20, 2020 of the data from Fig. 1 (Germany, left side) and Fig. 2 (US, right side) in a linear representation.**

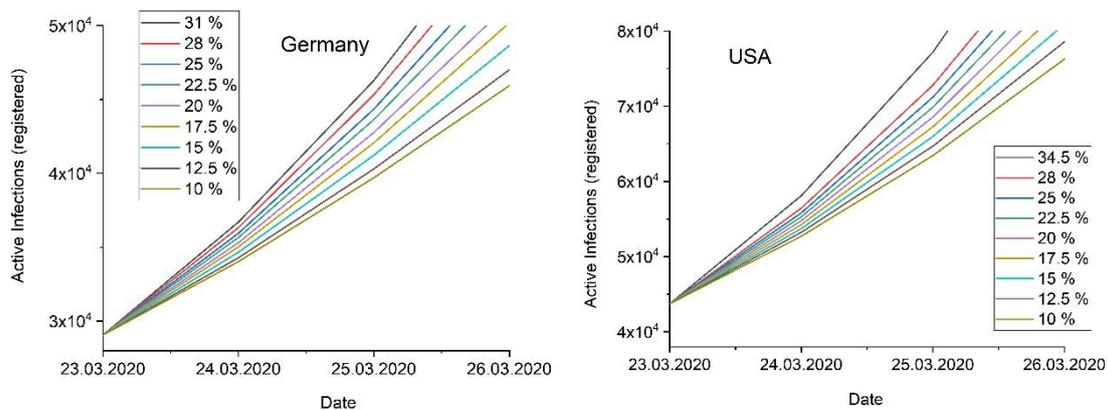

**Fig. 4 Enlargement of the period March 23, 2020 - March 26, 2020 of the data from Fig. 1 (Germany, left side) and Fig 2 (US, right side) in a linear representation.**

Compared to the literature, this model may seem extremely simplified. However, it turns out that essential statements found in the literature can be reproduced by this simple model (see also [4]).

**Acknowledgement**

Special thanks to my friends Zuleyha Yenice Campbell, Justus Füsers, Adam Tetzlaff, Claudia and Moritz Grehn, Marco Vitali, Markus Müller, Kinga Gerech, Marie Golüke, Michael Öchsner, Philipp Walk, Nikolas Pomplun, Jörn Weissenborn, Tom Küstner, Patrick Hätti and Jens Sobisch
as well as my colleagues Reinhard Krause-Rehberg, Mathias Stölzer, Nicki Hinsche and Florian Deininger for the motivation to deal with the topic presented here.